\documentclass[11pt]{article}

  \usepackage{latexsym}
  \usepackage{amssymb}

  \def\AFOUR{%
  \setlength{\textheight}{9.0in}%
  \setlength{\textwidth}{5.75in}

  \setlength{\topmargin}{-0.375in}%
  \hoffset=-.5in%
  \renewcommand{\baselinestretch}{1.17}%
  \setlength{\parskip}{6pt plus 2pt}%
  }

  \AFOUR                                          

  \makeatletter
  \def\section{\@startsection {section}{1}{\z@}{-3.5ex plus -1ex minus
  -.2ex}{2.3ex plus .2ex}{\large\bf}}
  \def\subsection{\@startsection{subsection}{2}{\z@}{-3.25ex plus -1ex minus
  -.2ex}{1.5ex plus .2ex}{\normalsize\bf}}
  \makeatother

  \makeatletter
  \@addtoreset{equation}{section}
  
  \makeatother
\newtheorem{theorem}{Theorem}[section]

\begin{document}
\begin{titlepage}

\begin{center}

\vspace{1cm}

\vskip .5in

{\Large\bf Hamiltonian formulation of the Grosse-Wulkenhaar
$\phi^{4}_{\star D}$ model} \vspace{10pt}

{\bf Mahouton Norbert Hounkonnou, Dine Ousmane Samary and Villevo
Adanhounme }

\vspace{10pt}

\vspace{0.5cm}

{\em University of Abomey-Calavi}\\
{\em International Chair of Mathematical Physics
and Applications} \\
{\em ICMPA-UNESCO Chair}\\
{\em 072 B.P. 50 Cotonou, Republic of Benin}\\
E-mail: norbert.hounkonnou@cipma.uac.bj\footnote{With copy to
hounkonnou@yahoo.fr}

\vspace{1.0cm}

\today

\begin{abstract}
A Hamiltonian formulation for the Grosse-Wulkenhaar
$\phi^{4}_{\star D}$ model is performed. The study is based on $D+1$ dimensional space-time formulation of $D$ dimensional non-local theories.
The analysis of
constraints shows that the secondary constraints describe the
Euler-Lagrange equations of motion. Relevant tensors are computed and analyzed.
\end{abstract}
\end{center}

\end{titlepage}
\makeatother
\section{Introduction}
The search of a unifying theory of gravity and quantum field
theory and the  obtaining of a better understanding of physics at
short distances have led to the development of the  noncommutative geometry.
 Subsequently, non-commutative field theories and
quantum gravity have been studied extensively. Such an approach  should lead to
change the nature of spacetime in a fundamental way. The  noncommutativity can be realized through the
coordinates which satisfy the commutation relations
$[\hat{x}^{\mu},\hat{x}^{\nu}]=i\Theta^{\mu\nu}(\hat{x})$.
$\Theta^{\mu\nu}(\hat{x})$ is unknown, but, for physical reasons,
should  vanish at large distances where we experience the
commutative world and may be determined by experiments. See
\cite{doplicher} and \cite{frank} and references therein. The algebra $\mathcal{M}$ of
functions of such noncommuting coordinates can be represented by
the algebra of functions on ordinary spacetime, equipped with a
noncommutative $\star-$product. A simple case of this  deformation
is the  D-dimensional Moyal space ${\rm I\!\!R}_\Theta^D$ endowed
with a constant Moyal $\star-$bracket of coordinate functions
\begin{eqnarray}
\left[x^\mu, x^\nu\right]_{\star} = i \Theta^{\mu\nu}
\end{eqnarray}
where $\Theta$ is a constant $D\times D$ non-degenerate
skew-symmetric matrix (which requires D even), usually  chosen in
the form
\begin{eqnarray}
\Theta=\theta J \mbox{ with } J= \left(\begin{array}{ll}
0 & I_{\frac{D}{2}}\\
-I_{\frac{D}{2}} & 0
\end{array}
\right).
\end{eqnarray}
$]0, +\infty[\ni\theta $ is a square length dimensional parameter,
($[\theta] = [L]^2$), $D$ denoting the spacetime dimension,
$I_{\frac{D}{2}}$ is the $D/2\times D/2$ identity matrix. The corresponding product of functions is the
associative, noncommutative Moyal-Groenewold-Weyl product, simply
called hereafter {Moyal product or $\star$-product} defined by
\begin{eqnarray}\label{9}
(f\star g)(x) = {\rm m}\left\{ e^{i \frac{\Theta^{\rho\sigma}}{2}
\partial_{\rho}\otimes \partial_{\sigma}} f(x)\otimes g(x)
\right\}\,\, \, x\in {\rm I\!\!R}_\Theta^D \qquad  \forall
f,g\in\mathcal{S}({\rm I\!\!R}_\Theta^D).
\end{eqnarray}
${\rm m}$ is the  ordinary multiplication of functions and
$\mathcal{S}({\rm I\!\!R}_\Theta^D)$ - the space of suitable
Schwartz functions.

The very process of replacing the point-wise multiplication of
functions at the same point by a star-product makes the theory
non-local. The star-product contains an infinite number of
space-time derivatives and this in turn affects the fundamental
causal structure on which all local, point-like quantum field
theories are built upon. Noncommutative field theories have
infinite number of space-time derivatives and then are non-local.
The non-local character of the theory also thanks to the property
that
\begin{eqnarray}
(f\star g)(x)& =&  \int d^Dy d^Dz \,K(x;y,z)f(y)g(z)
\end{eqnarray}
is evaluated through a two-point kernel
$ K(x;y,z)=\delta(x-y)\star\delta(x-z)
={e^{i(x-y)\Theta^{-1}(y-z)}\over{(2\pi)^D det\Theta}}
$. 
These theories have peculiar properties due to their acausal
behavior and lack of unitarity. The lack of unitarity is due to
the fact that $\Theta^{0i}\neq 0, i=1,2\cdots$.

This work is devoted to the construction of a Hamiltonian formulation 
of the Grosse-Wulkenhaar model,  one 
of the very few renormalizable noncommutative theories.
The Hamiltonian formulation of classical field theory, crucial in
the quantization procedure, remains a task to be solved in the
noncommutative field theories (NCFTs) widely developed in recent
years \cite{doplicher}-\cite{{Gracia-Bondia}} (and references
therein). So far, all attempts  to solve this problem have been
made before the advent of the new class of renormalizable NCFTs
built on the Grosse and Wulkenhaar (GW) $\phi^4$ scalar
field theory. See \cite{gommi} and \cite{llosa}
(and references therein) for more details. 
This paper aims at filling this gap, considering the class of
renormalizable  GW models  treated with a method that generalizes
previous construction \cite{gommi}. The expression of the total
Hamiltonian of the system is given. From a space-time Galilean
transformation and imposing an additional constraint from the
application of the Noether's theorem, the  Noether currents are
computed as a $U_{\star}(N)$ gauge currents from  $U_\star(N)$
gauge transformations with finite translations, i.e.
$g_\epsilon(x)=e^{-i\epsilon^\mu\Theta^{-1}_{\mu\nu}x^\nu}\in
U_\star(N) $ such that $g_\epsilon(x)\star f(x)\star
g_\epsilon^\dag(x)=f(x+\epsilon)$. The rotation group of $\mathbb{R}^D$ can be considered
as a particular concrete case.

\section{Hamiltonian formulation of  the NCFTs}
In this section, we briefly review the Hamiltonian formulation of NCFTs recently developed by
Gomis et al \cite{gommi}-\cite{llosa}. We then generalize this formulation by introducing a compact support function $w_h(x)$.
\subsection{Quick review of Hamiltonian formulation of NCFTs}
This subsection, mainly based on \cite{gommi} and \cite{llosa}
(and references therein), addresses a  Hamiltonian formulation of
field theories  in a noncommutative space-time. This formulation
involves two time coordinates $t$ and $\lambda$, and the dynamics
in this space is described in such a way that the evolution is
local with respect to one of the times. The non-local Lagrangian
at time $t$,  $L^{non}(t)$, depends not only on variables at time $t$  but also on
ones at different times. In other words, it depends on an infinite
number of time derivatives of the position $q_i(t)$. The analogue
of the tangent bundle for Lagrangians depending on positions and
velocities is now infinite dimensional and can be represented as
the space of all possible trajectories. The action is given by
\begin{eqnarray}\label{act}
S[q]=\int\, dt \, L^{non}(t)=\int\, dt \, L([q(t+\lambda)]).
\end{eqnarray}
The functional variational principle can be applied to the action
(\ref{act})  to produce the Euler-Lagrange (EL) equation of motion
as follows:
\begin{eqnarray}\label{cs}
\frac{\delta S[q]}{\delta q(t)}=\int\, dt'\,\frac{\delta
L^{non}(t')}{\delta q(t)}=0
\end{eqnarray}
which must be understood as a functional
relation to be fulfilled by physical trajectories. The latters are
not obtained as evolution of some given initial conditions. Since
the equation of motion is of infinite degree in time derivatives,
one should give as initial conditions the value of all these
derivatives at some initial time. In other words, we should give
the whole trajectory (or part of it) as the initial condition. Let
$J=\{q(\lambda), \lambda\in \mathbb{R}\}$ be the space of all
possible trajectories. Then the EL equation of motion  (\ref{cs})
is a Lagrangian constraint defining the subspace $J_R\subset J$ of
physical trajectories. In $1+1$ dimensional field theory, we
introduce new dynamical variables $\mathcal{Q}(t,\lambda)$ such
that
\begin{eqnarray}
\mathcal{Q}(t,\lambda)=q(t+\lambda)=T_tq(\lambda)
\end{eqnarray}
where $T_t$ is the time evolution operator for a given initial
trajectory $q(\lambda)$. $t$ is the evolution parameter and
$\lambda$ is a continuous parameter indexing the degrees of
freedom. If we denote by $\mathcal{P}(t,\lambda)$  the canonical
momentum of $\mathcal{Q}(t,\lambda)$, then the Hamiltonian is
defined as
\begin{eqnarray}
H(t,[\mathcal{Q},\mathcal{P}])=\int\, d\lambda\,
\mathcal{P}(t,\lambda)\mathcal{Q}'(t,\lambda)-\widetilde{L}(t,[\mathcal{Q}])
\end{eqnarray}
where
$\mathcal{Q}'(t,\lambda)=\partial_\lambda\mathcal{Q}(t,\lambda)$
and  $\widetilde{L}(t,[\mathcal{Q}])$ is a functional defined by
\begin{eqnarray}
\widetilde{L}(t,[\mathcal{Q}])=\int\,
d\lambda\,\delta(\lambda)\mathcal{L}(t,\lambda).
\end{eqnarray}
The density $\mathcal{L}(t,\lambda)$ is constructed from the
original non-local Lagrangian density $L^{non}(t)$ by replacing
$q(t)$ by $\mathcal{Q}(t,\lambda)$, the $t$-derivatives of $q(t)$
by $\lambda$-derivatives of $\mathcal{Q}(t,\lambda)$ and
$q(t+\rho)$ by $\mathcal{Q}(t,\lambda+\rho)$. In this construction
of the Hamiltonian,  $\lambda$ inherits the signature of the
original time $t$ and is a time-like coordinate.
$\mathcal{L}(t,\lambda)$ is local in $t$ and is non-local in
$\lambda$. $H$ depends linearly on $\mathcal{P}(t,\lambda)$ but
does not depend on $\dot{\mathcal{Q}}(t,\lambda)$.

The first and second Hamilton equations can be written as:
\begin{eqnarray}
\dot{\mathcal{Q}}(t,\lambda)=\mathcal{Q}'(t,\lambda),\,\,\dot{\mathcal{P}}(t,\lambda)
=\mathcal{P}'(t,\lambda)+\frac{\delta\widetilde{L}(t,[\mathcal{Q}])}{\delta
\mathcal{Q}(t,\lambda)}
\end{eqnarray}
where
$\dot{\mathcal{Q}}(t,\lambda)=\partial_t\mathcal{Q}(t,\lambda)$.
Their solutions  are related to those of the EL equations of
motion of the original non-local Lagrangian $L^{non}$ if we impose
a constraint on the momentum
\begin{eqnarray}
\varphi(t,\lambda)\equiv\mathcal{P}(t,\lambda)-\int\,d\sigma\,\frac{\epsilon(\lambda)
-\epsilon(\sigma)}{2}\frac{\delta\mathcal{L}(t,\sigma)}{\delta\mathcal{Q}(t,\lambda)}\approx
0.
\end{eqnarray}
Here $\epsilon(\lambda)$ is the sign distribution. The symbols
"$\equiv$" and "$\approx$ "  stand for "strong" and " weak"
equalities, respectively. Further constraints are generated by
requiring the stability of the primary ones. In the first step, we
obtain:
\begin{eqnarray}
\dot{\varphi}(t,\lambda)\equiv
\varphi'(t,\lambda)+\delta(\lambda)\int\,
d\sigma\,\frac{\delta\mathcal{L}(t,\sigma)}{\delta\mathcal{Q}(t,0)}\approx
0
\end{eqnarray}
or simply
\begin{eqnarray}
\dot{\varphi}_0(t,\lambda)\equiv\delta(\lambda)\int\,
d\sigma\,\frac{\delta\mathcal{L}(t,\sigma)}{\delta\mathcal{Q}(t,\lambda)}\approx
0
\end{eqnarray}
which reduces to the EL equation of motion. Repeating this, we get
an infinite set of Hamiltonian constraints. So doing, we are able
to describe the original non-local Lagrangian system as a $1+1$
dimensional local (in one of the times) Hamiltonian system,
governed by the Hamiltonian $H$ and a set of constraints. Note that
this formalism can be viewed as a generalization of the
Ostrogradski construction in the case of infinite order derivative
theories.
\subsection{A generalization in  $1+1$ dimensional field theory }
In this subsection, we aim at enlarging the class of Hamiltonians
that can be constructed in the framework of the above mentioned
formalism. The corresponding system of Hamilton equations and the
constraints are deduced.

Consider a parameter $h \in ]0,1[,$  $\quad x,\; y \in \mathbb{
R}^n$ and define
\begin{eqnarray}
|x-y| = \sqrt {\sum _{i=1}^{n}(x_{i}-y_{i})^2 },\,\,\;\; \omega
_{h}(x-y) = \frac{\omega (\frac{x-y}{h})}{h^n}
\end{eqnarray}
where
\begin{eqnarray}
\omega (u)= \left\{
\begin{array}{ll}
c\cdot\exp(\frac{1}{|u|^2-1})&  \quad |u|< 1\\
0 & \quad |u|\geq 1
\end{array}\right. , \,\,\,
c =\Big[\int _{|u|< 1}\exp(\frac{1}{|u|^2-1})du\Big]^{-1}.
\end{eqnarray}

Then we consider the family of Hamiltonians
\begin{eqnarray}\label{6}
\mathcal{H}_{h}(t,[\mathcal{Q}_{h},\mathcal{P}_{h}])=\int\,d\lambda
\mathcal{P}_{h}(t,\lambda)\mathcal{Q}'_{h}(t,\lambda)-L_{h}(t,[\mathcal{Q}_{h}])
\end{eqnarray}
where the quantities $\mathcal P_{h}(t,\lambda)$,    $\mathcal
Q_{h}(t,\lambda)$ ,    $L_{h}(t,[\mathcal{Q}_{h}])$ are defined as
follows:
\begin{eqnarray}
&&\mathcal P_{h}(t,\lambda)=\int _{\mathbb{ R}^2} dy \,\omega
_{h}(x-y)\mathcal P(y),\,\,
\mathcal Q_{h}(t,\lambda) = \int _ {\mathbb{ R}^2} dy \,\omega _{h}(x-y)\mathcal Q(y)  \\
&&L_{h}(t,[\mathcal Q_{h}]) =\int _{\mathbb{ R}^2} dy \,\omega
_{h}(x-y) L(t',\mathcal[Q])
\end{eqnarray}
$x=(t,\lambda)$ and $y=(t',\lambda ')$,
$\mathcal{Q}'_h(t,\lambda)=\partial_{\lambda}\mathcal{Q}_h(t,\lambda)$
and $\int _{\mathbb{ R}^n}dy\,\omega _{h}(x-y)=1$.

{\bf {Lemma}}: {\it  Let $L_{h},\quad h \in ]0,1[$, define a class
of  differentiable functionals  with compact support, i.e. for
$|x|<M,\quad L_{h}(x)\neq 0$, and $L_{h}(x)=0$ otherwise, where $M$
is a positive number:
\begin{eqnarray*}
L_{h}(x)=\int _{\mathbb{ R}^n} dy \omega _{h}(x-y)L(y).
\end{eqnarray*}
If the functional $L$ is summable on $x\in \mathcal{
D}\subset\mathbb{R}^{n}$, then
\begin{eqnarray}
\int _{\mathcal{ D}}\,|L_{h}(x)-L(x)|dx\rightarrow 0 \mbox{ for }
h\rightarrow 0,
\end{eqnarray}
i.e. $L_h$ converges in average to $L$  for $h\rightarrow 0$ where
$\mathcal{ D}$ is a bounded measurable set}.

{\bf {Proof}:} For  $x\in \mathcal{ D}$, we have
\begin{eqnarray}
L_{h}(x)-L(x)=\int _{\mathbb{R}^{n}}\, \omega
_{h}(x-y)[L(y)-L(x)]\mbox{d}y
\end{eqnarray}
where
$$
L(x)=\int _{\mathbb{R}^{n}} \omega
_{h}(x-y)L(x)\mbox{d}y=\frac{1}{h^{n}}\int _{|x-y|\leq
h}\,\omega\Big(\frac{x-y}{h}\Big)L(x)dy.
$$
Then
\begin{eqnarray}
&|L_{h}(x) - L(x)|\leq \frac{c_{1}}{h^{n}}\int _ {|x-y|\leq h}
|L(y)-L(x)|dy,
\\
&\mbox{ with }\qquad c_{1}=\max_{|x-y|\leq
h}\omega\Big(\frac{x-y}{h}\Big).
\end{eqnarray}
Using Fubini's theorem we can get
\begin{eqnarray}
\int _{\mathcal{ D}}|L_{h}(x)-L(x)|dx \leq\frac{c_1}{h^{n}}\int _{\mathcal{ D}}\int _{|x-y|\leq h} |L(y)-L(x)|dydx\\
\leq \frac{c_1}{h^{n}}\int_{|y'|\leq h}\mbox{d}y'\,\int
_{\mathcal{ D}} |L(x+y')-L(x)|\mbox{d}x.\nonumber
\end{eqnarray}
By the mean-continuity property, i.e. for all small $\epsilon
>0$ there exists a small $\delta >0$ such that
\begin{eqnarray}
\int _{\mathcal{ D}} |L(x+y')-L(x)|\mbox{d}x\leq \epsilon
,\textrm{ for } |y'|\leq \delta,
\end{eqnarray}
we obtain
\begin{eqnarray}
&&\int _{\mathcal{ D}}\,|L_{h}(x)-L(x)|\mbox{d}x \leq
\frac{c_1\epsilon}{h^{n}}\int _{|y'|\leq h}dy' \Rightarrow \int
_{\mathcal{ D}} |L_{h}(x)-L(x)|\mbox{d}x\leq c_1\epsilon\int
_{|z|\leq 1}\, \mbox{d}z \cr &&\Rightarrow\int _{\mathcal{ D}}
|L_{h}(x)-L(x)|\mbox{d}x\leq c_1\cdot c_2\epsilon\nonumber
\end{eqnarray}
where $c_2$ is the volume of the unit sphere in $\mathbb{R}^{n}$.
$\square$

Consider now the Lagrangian density $\mathcal{L}_{h}(t,\lambda)$
defined  by
\begin{eqnarray}
\mathcal{L}_{h}(t,[\mathcal Q_{h}])=\int_{\mathbb{ R}^2}\,dy\;
\omega(x-y)\mathcal{L}_{h}(y).
\end{eqnarray}
Following \cite{gommi}, the density $\mathcal L_{h}(t,\lambda)$ is
constructed from $\mathcal L_{h}^{non}(t)=L([q_{h}(t+\lambda)])$
by replacing $q_{h}(t)$ by $\mathcal Q_{h}(t,\lambda)$, the
$t$-derivatives of $q_{h}(t)$ by $\lambda$-derivatives of
$\mathcal Q_{h}(t,\lambda)$ and $q_{h}(t+\rho)$ by $\mathcal
Q_{h}(t,\lambda +\rho)$.  $\mathcal L_{h}(t,\lambda)$ is local in
t and nonlocal in $\lambda$. Defining then  a  time evolution operator
$T_{t}$ for a given initial trajectory $q(t)$  as follows
\begin{eqnarray}
T_{t}: q(\lambda)\mapsto q(t+\lambda),
\end{eqnarray}
we introduce a family of new dynamical variables $\mathcal
Q_{h}(t,\lambda)$, for $0<h<1$ as:
\begin{eqnarray}\label{7}
\mathcal Q_{h}(t,\lambda)=q_{h}(t+\lambda)=:T_t
\Big(q_h(\lambda)\Big).
\end{eqnarray}
 $t$ is the "evolution" parameter and $\lambda$  is a
continuous parameter indexing the degrees of freedom. In
differential form, condition (\ref{7}) reads:
\begin{eqnarray}\label{8}
\frac{\partial\mathcal Q_{h}}{\partial
t}(t,\lambda)=\frac{\partial\mathcal Q_{h}}{\partial
\lambda}(t,\lambda).
\end{eqnarray}
The fundamental Poisson bracket turns to be:
\begin{eqnarray}
\{\mathcal Q_{h}(t,\lambda),\mathcal P_{h}(t,\lambda ')\}
=\omega_{h}(\lambda -\lambda ').
\end{eqnarray}
The relation (\ref{8}) defines a family of first Hamilton
equations for (\ref{6}). The corresponding family of second
Hamilton equations can be written as follows:
\begin{eqnarray}
\dot{\mathcal P_{h}}(t,\alpha)&=& \mathcal
P'_{h}(t,\alpha)+\frac{\partial L_{h}(t,[\mathcal
Q_{h}])}{\partial\mathcal Q_{h}(t,\alpha)}
\end{eqnarray}
where $\mathcal P_{h}(t,\lambda)\omega_h(\lambda -\alpha)\Big
|_{\lambda=-M}^{M}=0$  ($\mathcal P_{h}$ with compact support).
Now integrating the second Hamilton equations yields
\begin{eqnarray}
\Gamma_{h}(t,\lambda ,[\mathcal Q_{h},\mathcal P_{h}]) \equiv
\mathcal P_{h}(t,\lambda)-\int d\sigma \frac{\delta\mathcal
L_{h}(t,\sigma)}{\delta\mathcal
Q_{h}(t,\lambda)}\cdot\frac{\epsilon(\lambda)
-\epsilon(\sigma)}{2}\approx 0.\nonumber\\
\end{eqnarray}
The stability of primary constraints implies the secondary
constraints given by
\begin{eqnarray}
\Xi_h\equiv\int\, d\lambda\,\frac{\delta
\mathcal{L}_{h}(t,\lambda)}{\delta \mathcal{Q}_{h}(t,0)} \approx
0.
\end{eqnarray}
\section{Hamiltonian formulation of
the Grosse-Wulkenhaar  model}
In this section, we first recall the GW model, derive the equation of motion and its solution in matrix base
and compute the Noether currents. Then we apply the generalized formulation of NC Hamiltonian construction developed in the previous 
section to the GW model and investigate the corresponding NC currents.
\subsection{ GW model}
Let us briefly recall the GW model and the Euler Lagrange
equations of motion with its solution in a matrix base
formalism. 

The renormalizable GW model is described by the
Lagrangian \cite{Grosse}
\begin{eqnarray}\label{lgw}
\mathcal L^{non}(t)&=
&\mathcal{L}_{\star}[\phi,\partial_{\mu}\phi]=\frac{1}{2}\partial
_{\mu}\phi(x)\star\partial ^{\mu}\phi(x)+\frac{\Omega
^{2}}{2}\Big(\tilde{x}_{\mu}
\phi(x)\Big)\star\Big(\tilde{x}^{\mu}\phi(x)\Big)\nonumber\\
&+&\frac{m^2}{2}\phi(x)\star\phi(x)
+\frac{\lambda}{4!}\phi(x)^{4}_{\star},
\end{eqnarray}
where $\tilde{x}_{\mu}=2(\Theta ^{-1})_{\mu\nu}x^{\nu}$ and
$\phi_{\star}^n=\underbrace{\phi\star\phi\cdots
\star\phi}_{\mbox{n times}}$. { $\Theta$} breaks into diagonal
blocks $\left(\begin{array}{ll}
0 & \theta \\
-\theta & 0
\end{array}\right)$. The harmonic term $\Omega$ ensures ultraviolet
(UV)/infrared (IR) freedom for the action implying its
renormalizability, and such that the Lagrangian action becomes
covariant under Langmann-Szabo duality  \cite{LS}, i.e. covariant
under the symmetry: $\tilde{x}_{\mu}\longleftrightarrow
p_{\mu}\equiv \partial_{\mu}$ giving

\begin{eqnarray}
\mathcal{S}[\phi, m, \lambda, \Omega]\rightarrow\Omega^2
\mathcal{S}[\phi, \frac{m}{\Omega}, \frac{\lambda}{\Omega^2},
\frac{1}{\Omega}].
\end{eqnarray}

The Lagrangian density depending explicitly on $x^\mu$, through
the field $\phi$  interaction with a harmonic external source,
does not describe a closed system.  Furthermore, it is not
invariant under space-time translation.  Besides, at the parameter
limit $\theta \to 0$, the model does not converge to the ordinary
$\phi^4$ scalar field theory due to the presence of the inverse
matrix $(\Theta^{-1})$, then causing a singularity. The  {
$\star$}-Grosse-Wulkenhaar { $\phi_D^4$}  theory is renormalizable
at all orders in $\lambda$. This result has been now proved by
various methods (see \cite{Rivasseau} and references therein).
For more details on the properties of this model, see
\cite{Grosse}, \cite{Ben-Houk1} and \cite{Ben-Houk2} (and
references therein). The $\star$-product of two real fields is not
necessarily real, and the non-locality comes from the infinite
derivatives in (\ref{9}). 

The peculiar EL equations of motion can be readily derived for the
field $\phi$ by direct application of the variational principle. We get 
\begin{eqnarray}
 \frac{\delta \mathcal{S}}{\delta\phi}=\int\, d^Dx\,\Big(-\partial _{\rho}\partial ^{\rho}\phi +m^{2}\phi
+\frac{\lambda}{3!}\phi ^{3}_{\star}+\Omega^2\,\tilde{x}\star\tilde{x}\star\phi\Big)=0\nonumber
\end{eqnarray}
which gives, using the identity $\int\,d^Dx \,(f\star g)=\frac{1}{2}\int\,d^Dx\,(f\star g+g\star f)$,
\begin{eqnarray}\label{motion}
\frac{\delta \mathcal{S}}{\delta\phi}=0\Longleftrightarrow
-\partial _{\rho}\partial ^{\rho}\phi +m^{2}\phi
+\frac{\lambda}{3!}\phi ^{3}_{\star}
+\frac{\Omega ^{2}}{4}\Big( 2\tilde{x}\star\phi\star\tilde{x}+\{\phi ,\tilde{x}\star\tilde{x}\}_{\star}\Big)=0,\nonumber\\
\end{eqnarray}
where $\{.,.\}_{\star}$ defines the star anticommutator.

 Remark that from the equation
\begin{eqnarray}
\frac{\delta \mathcal{S}}{\delta\tilde{x}_\nu}=0 \Leftrightarrow
\Omega^2(2\phi\star \tilde{x}^\nu\star\phi+\tilde{x}^\nu\star
\phi_{\star}^2+\phi_{\star}^2\star\tilde{x}^\nu
)+\Theta^{\mu\nu}\partial_{\mu}\phi\star \frac{\delta
\mathcal{S}}{\delta\phi}=0
\end{eqnarray}
we get the additional constraint
\begin{eqnarray}\label{constr}
\pi(\phi,\tilde{x} )=\Omega^2(2\phi\star
\tilde{x}^\nu\star\phi+\tilde{x}^\nu\star
\phi_{\star}^2+\phi_{\star}^2\star\tilde{x}^\nu )\approx0.
\end{eqnarray}
de Goursac {\it et al} \cite{Goursac}  solved the equation of motion (\ref{motion}),  
representing the elements on the $D-$dimensional Moyal algebra
$\mathcal{M}$ with the help of a matrix base
\cite{Gracia-Bondia} whose elements $b_{kl}^{(D)}(x)$ are
eigenfunctions of the harmonic oscillator Hamiltonian
$$H=\sum_{l=1}^{\frac{D}{2}}\frac{1}{2}\Big(x_{2l-1}^{2}+x_{2l}^{2}\Big),$$ with
$b_{00}^{(D)}=2^{D/2}e^{-2H/\theta}$ such that $b_{00}^{(D)}\star
b_{00}^{(D)}=b_{00}^{(D)}$. Then defining the operators
$$
a_{l}=\frac{x_{2l-1}+ix_{2l}}{\sqrt{2}},\mbox{ and }
\bar{a}_{l}=\frac{x_{2l-1}-ix_{2l}}{\sqrt{2}}
$$
together with
\begin{eqnarray}
b_{kl}^{(D)}=\frac{\bar{a}_{\star}^{k}\star b_{00}^{(D)}\star
a_{\star}^{l}}{\sqrt{k!l!\theta^{|k|+|l|}}}
\end{eqnarray}
where $a=\sum_{i=1}^{D/2}a_{i}$, and
$\bar{a}=\sum_{i=1}^{D/2}\bar{a}_{i}$, one has the following
left and right actions:
\begin{eqnarray}
&&a\star b_{kl}^{(D)}=\sqrt{|k|\theta}b_{k-1,l}^{(D)},\quad b_{kl}^{(D)}\star a= \sqrt{\theta(|l+1|)}b_{k,l+1}^{(D)},\\
&&\overline{a}\star
b_{kl}^{(D)}=\sqrt{\theta(|k+1|)}b_{k+1,l}^{(D)},\quad
b_{kl}^{(D)}\star \overline{a}=\sqrt{|l|\theta}b_{k,l-1}^{(D)}
\end{eqnarray}
and
\begin{eqnarray}
H\star b_{kl}^{(D)}=\theta(|k|+\frac{1}{2})b_{kl}^{(D)},\quad
b_{kl}^{(D)}\star H=\theta(|l|+\frac{1}{2})b_{kl}^{(D)},
\end{eqnarray}
where $k, l\in \mathbb{N}^{D/2}$ and $|k|=\sum_{i=1}^{D/2}k_{i}$.
For $D=2$, $b_{kl}^{(2)}=f_{kl}$ which can be expanded in polar
coordinates,  ($x_{1}=rcos (\varphi),\quad x_{2}=rsin (\varphi)$),
to give
\begin{eqnarray}
f_{kl}=2(-1)^{k}\sqrt{\frac{k!}{l!}}e^{i(l-k)\varphi}\Big(\frac{2r^{2}}{\theta}\Big)^{\frac{l-k}{2}}L_{k}^{l-k}\Big(\frac{2r^{2}}{\theta}\Big)
e^{-\frac{r^{2}}{\theta}}
\end{eqnarray}
where the $L_{n}^{k}(x)$ are the associated Laguerre polynomials.
The generalization to higher dimensions  is straightforward. In
particular, for $D=4$, one gets $k=(k_1, k_2)$, $l=(l_1 ,l_2)$ and
$$
b_{kl}^{(4)}(x)=f_{k_{1},l_{1}}(x_{1},x_{2})f_{k_{2},l_{2}}(x_{3},x_{4}).
$$
More generally,  the following properties are satisfied:
\begin{eqnarray}
(b_{kl}^{(D)}\star b_{k'l'}^{(D)}) (x)&=&\delta_{lk'}b_{kl'}^{(D)}(x),\\
\int\, d^{D}x\,\,\, b_{kl}^{(D)}(x)&=&(2\pi\theta)^{D/2}\delta_{kl},\\
(b_{kl}^{(D)})^{\dag}&=&b_{lk}^{(D)}.
\end{eqnarray}
The existence of an isomorphism between the unital involutive
Moyal algebra and a subalgebra of the unital involutive algebra of
complex infinite-dimensional matrices allows to define, for all
$g\in \mathcal{M}$,  a unique matrix $(g_{kl})$ given by
$$
g_{kl}=\frac{1}{(2\pi\theta)^{D/2}}\int \, d^{D}x
\,\,\,g(x)b_{kl}^{(D)}
$$
satisfying
$$
\forall x\in \mathbb{R}^{D},\quad
g(x)=\sum_{k,l\in\mathbb{N}^{D/2}}g_{kl}\,b_{kl}^{(D)}(x).
$$
Setting $\phi(x)=\tau b_{kl}^{(D)}(x)$, where $\tau$ is a complex
constant, considering
\begin{eqnarray}
&&\tilde{x}\star\tilde{x}\star b_{kl}^{(D)}=-\frac{8}{\theta^{2}}
H\star b_{kl}^{(D)}=-\frac{8}{\theta}(|k|+\frac{1}{2})
b_{kl}^{(D)},
\end{eqnarray}
and
\begin{eqnarray}
\tilde{x}\star b_{kl}^{(D)}\star
\tilde{x}&=&-\frac{4}{\theta^{2}}(a\star b_{kl}^{(D)}
\star \bar{a}+\bar{a}\star b_{kl}^{(D)}\star a)\\
&=&-\frac{4}{\theta}\Big(\sqrt{|k||l|}b_{k-1,l-1}^{(D)}+\sqrt{|k+1||l+1|}b_{k+1,l+1}^{(D)}\Big),
\end{eqnarray}
and taking into account the relations
$$
[\tilde{x}_{\mu},\phi]_{\star}=2i\partial_{\mu}\phi\Rightarrow
\partial_{\mu}\partial^{\mu}\phi=-\frac{1}{4}[\tilde{x}_{\mu},[\tilde{x}_{\mu},\phi]_{\star}]_{\star},
$$
the equation of motion (\ref{motion}) can be rewritten in the form
\begin{eqnarray}
\frac{1}{2}(\Omega^2-1)\tilde{x}\star\phi\star\tilde{x}+\frac{1}{4}(\Omega^2+1)\Big(\tilde{x}
\star\tilde{x}\star\phi+\phi\star\tilde{x}\star\tilde{x}\Big)+m^2\phi+\frac{\lambda}{3!}\phi^3_\star=0
\end{eqnarray}
or equivalently
\begin{eqnarray}\label{motion2}
&&-\frac{2}{\theta}(\Omega^2-1)\Big(\sqrt{|k||l|}b_{k-1,l-1}^{(D)}+\sqrt{|k+1||l+1|}b_{k+1,l+1}^{(D)}\Big)\cr
&&-\frac{2}{\theta}(\Omega^2+1)\Big(|k|+|l|+|1|\Big)b_{kl}^{(D)}+\Big(m^2+\frac{\lambda}{3!}\tau^2\Big)b_{kl}^{(D)}=0.
\end{eqnarray}
If $\Omega=1$, then (\ref{motion2}) is reduced to
\begin{eqnarray}
\Big[-\frac{4}{\theta}\Big(|k|+|l|+|1|\Big)+\Big(m^2+\frac{\lambda}{3!}\tau^2\Big)\Big]b_{kl}^{(D)}=0.
\end{eqnarray}
The vectors $k$ and $l$ can be chosen such that
$\frac{4}{\theta}\Big(|k|+|l|+|1|\Big)\geq m^2$. In this case, the suitable solution is
$\tau=\Big(\frac{3!}{\lambda}\Big)^{1/2}\Big[\frac{4}{\theta}\Big(|k|+|l|+|1|\Big)-
m^2\Big]^{1/2}$ and finally  
\begin{eqnarray}
\phi(x)=\Big(\frac{3!}{\lambda}\Big)^{1/2}\Big[\frac{4}{\theta}\Big(|k|+|l|+|1|\Big)-
m^2\Big]^{1/2}b_{kl}^{(D)}(x).
\end{eqnarray}
More details can be found in \cite{Goursac}.
\subsection{Noether currents}
Let us now consider a Lie group of continuous transformations
$$x^\mu\longmapsto x'^\mu=x'^\mu(x)$$
$$\phi(x)\longmapsto
\phi'(x')=\phi'(\phi(x),x)$$ inducing a set of infinitesimal
transformations
\begin{eqnarray}\label{tran1}
\delta x^\mu
(\omega,\epsilon)=\omega_{\nu}^{\mu}x^\nu+\epsilon^\mu,\,\,\,
\omega_{\mu\nu}=-\omega_{\nu\mu}\\\label{tran2}
\delta_\pm\phi(\omega,\epsilon,\mathcal{X},\xi)=\mathcal{X}\star\phi+\phi\star\xi+(\delta
x^\mu\star\partial_{\mu}\phi)_{\pm},
\end{eqnarray}
where
$\omega_{\mu\nu}$ is an antisymmetric constant matrix,
$\epsilon^\mu$ a constant vector. $\mathcal{X}=\mathcal{X}(x)$ and
$\xi=\xi(x)$ are two families of functions, spanning the Lie
algebra of the Lie group of fixed
dimension $r$; \Big($(f\star g)_{+}=f\star g$, $(f\star g)_{-}=g\star f$\Big).
 The transformations (\ref{tran1}) and (\ref{tran2}) actually meet
known infinitesimal transformations in NCFTs. Indeed, generators
of deformed Poincar\'e or Galilean transformations are recovered
for $\mathcal{X}=0=\xi$.
\begin{eqnarray}\label{six}
\delta_\pm(.)=(\delta x^\mu\star\partial_{\mu}(.))_{\pm},\,\,\,
\delta_{\pm}\phi=(\omega_\nu^\mu
x^\nu\star\partial_{\mu})_\pm+\epsilon^\mu\partial_\mu\phi.
\end{eqnarray}
In this situation,  a $\star-$deformed Poincar\'e or Galilean
algebra can be  defined by the generators
$p_\mu(.)=\partial_{\mu}(.)$ and
$m_{\pm}(\omega)(.)=(\omega_\nu^\mu
x^\nu\star\partial_{\mu}(.))_\pm$ satisfying
\begin{eqnarray}
&&[p_\mu,m_{\pm}(\omega)](.)=\omega_{\mu}^\nu p_\nu,\\
&&[m_{\pm}(\omega),m_{\pm}(\omega')](.)=(\omega\times\omega')_\rho^\nu(x^\rho\star\partial_{\nu}(.))_{\pm}=\delta_{\pm}(\omega\times\omega')(.)\\
&&(\omega\times\omega')_{\rho}^{\nu}:=-(\omega_{\mu}^{\nu}\omega^{'\mu}_{\rho}-\omega_{\mu}^{'\nu}\omega_{\rho}^{\mu}).
\end{eqnarray}
Besides, pure translation symmetry is obviously obtained by
setting $\omega=0$ in (\ref{six}). A finite dimensional group of
transformations can be obtained by simple exponentiation of these
infinitesimal generators. From the infinitesimal transformation
(\ref{six}), we can now define the generalized global Ward
identity operator (WIop) as follows \cite{Gerh}, \cite{Ben-Houk1}:
\begin{eqnarray}
\mathcal{W}&=&\frac{1}{2}\int d^{D}x\,(\delta\phi\star
\frac{\delta(.)}{\delta\phi}+\frac{\delta(.)}{\delta\phi}\star\delta\phi
+\delta\tilde{x}_{\rho}
\star\frac{\delta(.)}{\delta\tilde{x}_{\rho}}+\frac{\delta(.)}{\delta\tilde{x}_{\rho}}\star\delta\tilde{x}_{\rho})
\end{eqnarray}
such that its action on the Lagrangian action gives, (after
lengthy but straightforward computations),
\begin{eqnarray}
\mathcal{W}\mathcal{S}&=&
\int\,d^{D}x\,\Big[-\epsilon^\mu\partial^{\rho}\mathcal{T}_{\rho\mu}
-\frac{\omega^{\mu\nu}}{2}\partial^{\rho}\Big(x_\nu\star\mathcal{T}_{\rho\mu}-x_\mu\star\mathcal{T}_{\rho\nu}\Big)+\mathcal{B}(\omega)\Big]
\end{eqnarray}
where the Galilean invariance breaking term $\mathcal{B}(\omega)$
is given by
\begin{eqnarray}
\mathcal{B}(\omega)&=&-\omega^{\mu\nu}x_\nu\star\Big(\frac{\lambda}{4!}[[\partial_{\mu}\phi,
\phi]_{\star},\phi\star\phi]_{\star}+\frac{\Omega^2}{8}[[\partial_{\mu}\phi,\{\tilde{x}_{\nu},
\phi\}_{\star}]_{\star},\tilde{x}^\nu]_{\star}\Big),
\end{eqnarray}
while the canonical energy momentum tensor $\mathcal{T}_{\rho\mu}$
and the broken angular momentum tensor $\mathcal{M}_{\nu\rho\mu}$
are expressed by the  relations
\begin{eqnarray}\label{tensor}
\mathcal{T}_{\rho\mu}=\frac{1}{2}\{\partial_{\rho}\phi,\partial_{\mu}\phi\}_{\star}
-g_{\rho\mu}\mathcal{L}_{\star},\,\,\,
\mathcal{M}_{\nu\rho\mu}=x_\nu\star\mathcal{T}_{\rho\mu}-x_\mu\star\mathcal{T}_{\rho\nu},
\end{eqnarray}
respectively. $g_{\rho\mu}$ stands for the Euclidean metric. In the
particular case where $\omega=0$ (pure translation), the action
becomes invariant and $\mathcal{W}\mathcal{S}=0$.
\subsection{Hamiltonian formulation of the GW model in $D+1$ dimensions }
We now consider  the transformation of the $D$ canonical field
variables into the $D+1$ ones
\begin{eqnarray*}
x^{\mu}=(t,x^i)\longmapsto
X^{\mu}=(t,x^{0},x^{i})=(t,\bar{x}^i),\,\, \phi(x)\longmapsto
\mathcal Q(t,\bar{x})
\end{eqnarray*}
and
$\tilde{x}\longmapsto\tilde{X}=(t,\tilde{\bar{x}})=(t,2(\Theta
^{-1})\bar{x})$. In this case,
  the  Lagrangian density takes the form
\begin{eqnarray}\label{10}
\mathcal L^{non}(t,\bar{x})&=& \frac{1}{2}\partial _{\mu}\mathcal
Q(t,\bar{x})\star\partial ^{\mu}\mathcal Q(t,\bar{x})+
\frac{\Omega ^{2}}{2}\Big(\tilde{\bar{x}}_{\mu}\mathcal Q(t,\bar{x})\Big)\star\Big(\tilde{\bar{x}}^{\mu}\mathcal Q(t,\bar{x})\Big)\nonumber\\
&+&\frac{m^2}{2}\mathcal Q(t,\bar{x})\star \mathcal Q(t,\bar{x})
+\frac{\lambda}{4!}\mathcal Q(t,\bar{x})\star \mathcal Q(t,\bar{x})\star \mathcal Q(t,\bar{x})\star \mathcal Q(t,\bar{x}).\nonumber\\
\end{eqnarray}
Substituting $  \phi(x) \quad \textrm {by} \quad \mathcal
Q_{h}(t,x^{0},x^{i}) $, we get the family of Lagrangian densities
\begin{eqnarray} \label{11}
\mathcal L^{non}_{h}(t,\bar{x}) &=&\frac{1}{2}\partial
_{\mu}\mathcal Q_{h}(t,\bar{x})\star \partial ^{\mu}\mathcal
Q_{h}(t,\bar{x})
+\frac{m^2}{2}\mathcal Q_{h}(t,\bar{x})\star\mathcal  Q_{h}(t,\bar{x}) \nonumber\\
&+&\frac{\Omega ^{2}}{2}\Big(\tilde{\bar{x}} _{\mu}\mathcal Q_{h}(t,\bar{x})\Big)\star\Big(\tilde{\bar{x}}^{\mu}\mathcal Q_{h}(t,\bar{x})\Big)\nonumber\\
&+& \frac{\lambda}{4!}\mathcal Q_{h}(t,\bar{x})\star\mathcal
Q_{h}(t,\bar{x})\star\mathcal Q_{h}(t,\bar{x})\star \mathcal
Q_{h}(t,\bar{x}).
\end{eqnarray}

In view of computing the constraints, let us define 
the symmetric Kernel $K$ of four star products by
\begin{eqnarray}
K(x-x_1 ,x-x_2 ,x-x_3 ,
x-x_4)=e^{-ix\wedge\sum_{i=1}^{4}(-1)^{i+1}x_{i}}e^{-i\varphi_{4}}
\end{eqnarray}
where $\varphi_{4}=\sum_{i<j=1}^{4}(-1)^{i+j+1}x_{i}\wedge x_{j}, $
$x\wedge y=2x\Theta^{-1}y.$
 In expanded form,  we
get
\begin{eqnarray}
&&K(x-x_1 ,x-x_2 ,x-x_3 ,
x-x_4)=\exp\Big\{-i\Big[(x-x_1)\wedge(x-x_2)\cr
&&-(x-x_1)\wedge(x-x_3)+(x-x_1)\wedge(x-x_4)+(x-x_2)\wedge(x-x_3)\cr
&&-(x-x_2)\wedge(x-x_4)+(x-x_3)\wedge(x-x_4)\Big]\Big\}.
\end{eqnarray}
Then the quantity
\begin{eqnarray}
\Upsilon _{h} (t,\bar{x}):&=&\int d^{D}\bar{x}'\frac{\delta\mathcal L_{h}(t,\bar{x}')}{\delta\mathcal Q_{h}(t,\bar{x})}\cdot\frac{\epsilon(\bar{x}^{0})-\epsilon(\bar{x}'^{0})}{2} \nonumber\\
&=& -\delta(\bar{x}^{0})\partial _{\bar{x}^{0}}\mathcal Q_{h}(t,\bar{x})+\frac{\lambda}{4!}\int d^{D}y_{1}d^{D}y_{2}d^{D}y_{3} d^{D}\bar{x}' \Big(\frac{\epsilon(\bar{x}^{0})-\epsilon(\bar{x}'^{0})}{2}\Big)\nonumber\\
&&\times\mathcal Q_{h}(t,y_{1})\mathcal Q_{h}(t,y_{2})\mathcal Q_{h}(t,y_{3})\Phi(x,y_1,y_2,y_3,y_4)\nonumber\\
&+&\frac{\Omega ^{2}}{8}\int d^{D}y_{1}d^{D}y_{2}d^{D}y_{3}d^{D}\bar{x}'\, \tilde{y}_{1}\tilde{y}_{2}\mathcal Q_{h}(t,y_{3})\frac{\epsilon(\bar{x}^{0})-\epsilon(\bar{x}^{'0})}{2}\nonumber\\
&&\times\Psi(x,y_1,y_2,y_3,y_4)
\end{eqnarray}
with
\begin{eqnarray}
\Phi(x,y_1,y_2,y_3,y_4)&=&\Big[K(\bar{x}-\bar{x}', y_1 -\bar{x}', y_2 -\bar{x}', y_3 -\bar{x}')\nonumber\\
&+& K(y_1 -\bar{x}',\bar{x}-\bar{x}', y_2 -\bar{x}', y_3 -\bar{x}')  \nonumber\\
&+&K( y_1 -\bar{x}', y_2 -\bar{x}',\bar{x}-\bar{x}', y_3 -\bar{x}')\nonumber\\
&+&K( y_1 -\bar{x}', y_2 -\bar{x}', y_3
-\bar{x}',\bar{x}-\bar{x}')\Big]
\end{eqnarray}
and
\begin{eqnarray}
\Psi(x,y_{1},y_{2},y_{3},y_{4})&=& \Big[ K(\bar{x}'-y_{1},\bar{x}'-\bar{x},\bar{x}'-y_{2},\bar{x}'-y_{3})\nonumber\\
&+& K(\bar{x}'-\bar{x},\bar{x}'-y_{1},\bar{x}'-y_{3},\bar{x}'-y_{2})\nonumber\\
&+& K(\bar{x}'-y_{1},\bar{x}'-y_{3},\bar{x}'-\bar{x},\bar{x}'-y_{2})\nonumber\\
&+& K(\bar{x}'-y_{2},\bar{x}'-\bar{x},\bar{x}'-y_{1},\bar{x}'-y_{3})\nonumber\\
&+& K(\bar{x}'-y_{1},\bar{x}'-y_{3},\bar{x}'-y_{2},\bar{x}'-\bar{x}) \nonumber\\
&+& K(\bar{x}'-y_{3},\bar{x}'-y_{1},\bar{x}'-\bar{x},\bar{x}'-y_{2})\nonumber\\
&+& K(\bar{x}'-y_{1},\bar{x}'-\bar{x},\bar{x}'-y_{3},\bar{x}'-y_{2}) \nonumber\\
&+&
K(\bar{x}'-y_{2},\bar{x}'-y_{3},\bar{x}'-y_{1},\bar{x}'-\bar{x})\Big],
\end{eqnarray}
allows to compute the family of primary constraints for the class of
GW models defined by the parameter $h$ as follows:
\begin{eqnarray}\label{13}
\Gamma _{h}(t,\bar{x})\equiv \mathcal P_{h}(t,\bar{x}) - \Upsilon
_{h}(t,\bar{x})\approx 0.
\end{eqnarray}
 The family of secondary
constraints can be obtained in the same way. The previous lemma
guarantees the convergence:
\begin{eqnarray}
\textrm {h}\rightarrow 0 \Rightarrow \mathcal P_{h} \rightarrow
\mathcal P;  \qquad    \mathcal Q_{h} \rightarrow \mathcal Q
;\qquad \Gamma _{h}    \rightarrow  \Gamma
\end{eqnarray}
as well as the limit of the family (\ref{13}) of primary
constraints:
\begin{eqnarray*}
\Gamma=\lim_{h\rightarrow 0}\Gamma_{h}=\mathcal P(t,\bar{x}) -
\Upsilon(t,\bar{x})\approx 0.
\end{eqnarray*}
The secondary constraints appear as the equation of motion of the
field $\mathcal{Q}_h$, i.e.
\begin{eqnarray}\label{bis13}
  \Xi_{h}(t,\bar{x})\approx 0.
\end{eqnarray}
 The total Hamiltonian can be then defined as
\begin{eqnarray}\label{tha}
&\mathcal{H}_{h}^{T}(t,[\mathcal{Q}_h,\mathcal{P}_h])=\mathcal{H}_{h}(t,[\mathcal{Q}_h,\mathcal{P}_h])+\Lambda^{1} (t,\bar{x})\star\Gamma_{h}(t,\bar{x})\nonumber\\
&+\Lambda^{2} (t,\bar{x})\star\Xi_{h}(t,\bar{x}),
\end{eqnarray}
where  $\Lambda ^{i}(t,\bar{x}), i=1,2$ are Lagrange multipliers.
The corresponding field theory action
$\mathcal{S}_{h}^{T}(t,\bar{x})$
\begin{eqnarray}
\mathcal{S}_{h}^{T}(t,\bar{x}) &=&    \int dt d^{D}\bar{x}\,\Big(\mathcal{L}_{h}(t,\bar{x})+\Lambda^{1} (t,\bar{x})\star\Gamma_{h}(t,\bar{x})+\Lambda^{2} (t,\bar{x})\star\Xi_{h}(t,\bar{x})\Big)\nonumber\\
&=&\int dt d^{D}\bar{x}\,\mathcal{L}_{h}^{T}(t,\bar{x}),\qquad
\Lambda^{i} (t,\bar{x})\in T^{*}J
\end{eqnarray}
generates the Euler-Lagrange equation of motion
\begin{eqnarray}
\frac{\delta \mathcal{S}_{h}^{T}(t,\bar{x})}{\delta \mathcal{Q}_{h}(t,\bar{x}')}&=&\int dtd^{D}\bar{x}\,\Big(\frac{\delta \mathcal{L}_{h}(t,\bar{x})}{\delta \mathcal{Q}_{h}(t,\bar{x}')}+\Lambda ^{1}(t,\bar{x})\star\frac{\delta \Gamma_{h} (t,\bar{x})}{\delta \mathcal{Q}_{h}(t,\bar{x}')}\nonumber\\
&+&\frac{\delta\Lambda^{1}(t,\bar{x})}{\delta\mathcal{Q}_{h}(t,\bar{x}')}\star\Gamma_{h}(t,\bar{x})+\Lambda ^{2}(t,\bar{x})\star\frac{\delta \Xi_{h} (t,\bar{x})}{\delta \mathcal{Q}_{h}(t,\bar{x}')}\nonumber\\
&+&\frac{\delta\Lambda^{2}(t,\bar{x})}{\delta\mathcal{Q}_{h}(t,\bar{x}')}\star\Xi_{h}(t,\bar{x})\Big)=0
\end{eqnarray}
which  gives
\begin{eqnarray}
\frac{\delta \mathcal{L}_{h}(t,\bar{x})}{\delta
\mathcal{Q}_{h}(t,\bar{x}')}+ \Lambda
^{1}(t,\bar{x})\star\frac{\delta \Gamma_{h}(t,\bar{x})}{\delta
\mathcal{Q}_{h}(t,\bar{x}')} +\Lambda
^{2}(t,\bar{x})\star\frac{\delta \Xi_{h} (t,\bar{x})}{\delta
\mathcal{Q}_{h}(t,\bar{x}')}\Big)\approx 0
\end{eqnarray}
where the constraints equations (\ref{13}) and (\ref{bis13}) have been taken into account. If we
perform the following set of infinitesimal transformations of
simply connected continuous arbitrary group $G$:
\begin{eqnarray}\label{14}
&&\bar{x}\longmapsto \bar{x}'=\bar{x}+\frac{1}{2}\Big(\varpi
_{a}\star\frac{\delta \bar{x}}{\delta\varpi _{a}}
+\frac{\delta \bar{x}}{\delta\varpi _{a}}\star\varpi _{a}\Big),\;\;({a=1,2,\cdots})\nonumber\\
&&\mathcal{Q}_{h}(t,\bar{x})\longmapsto
\mathcal{Q}^{t}_{h}(t,\bar{x}')=
\mathcal{Q}_{h}(t,\bar{x})+\frac{1}{2}\Big(\varpi
_{a}\star\frac{\delta
\mathcal{F}(\mathcal{Q}_{h}(t,\bar{x}))}{\delta \varpi _{a}}+
\frac{\delta \mathcal{F}(\mathcal{Q}_{h}
(t,\bar{x}))}{\delta \varpi _{a}}\star\varpi _{a}\Big),\nonumber\\
\end{eqnarray}
where  $\mathcal{F}(\mathcal{Q}_{h}(t,\bar{x}))$ is a
transformation of fields $\mathcal{Q}_{h}(t,\bar{x})$ and
$\{\varpi _{a}(\bar{x})\}$ defines a family of infinitesimal
parameters of this group, then the transformation
$\mathcal{Q}^{t}_{h}(t,\bar{x}')$ of fields
$\mathcal{Q}_{h}(t,\bar{x}')$ at a same point $\bar{x}'$ can be
expressed through the generators $G_\mu^{a}$ as:
\begin{eqnarray}
\mathcal{Q}^{t}_{h}(t,\bar{x}')=\Big(1-\frac{i}{2}\{\varpi _{a},
G_\mu^{a}\}_{\star} +\textbf{O}(\varpi^{2})\Big)\star
\mathcal{Q}_{h}(t,\bar{x}')=e_{\star}^{-\frac{i}{2}\{\varpi _{a},
G_\mu^{a}\}_{\star}}\star \mathcal{Q}_{h}(t,\bar{x}'),
\end{eqnarray}
with
\begin{eqnarray}
e^{i\alpha}_{\star}=1 + i\alpha +\frac{i^2}{2!} \alpha\star\alpha
+ \frac{i^3}{3!} \alpha\star\alpha \star\alpha +\dots; \alpha \in
C^\infty({\mathbb{ R}})
\end{eqnarray}
  and
\begin{eqnarray}
\mathcal{Q}_{h}(t,\bar{x}')&=&\mathcal{Q}_{h}\Big(t,\bar{x}+\frac{1}{2}(\varpi
_{a}\star\frac{\delta \bar{x}}{\delta\varpi _{a}}+\frac{\delta
\bar{x}}{\delta\varpi _{a}}\star\varpi _{a})\Big)\cr
&=&\mathcal{Q}_{h}(t,\bar{x})+\frac{1}{2}(\varpi
_{a}\star\frac{\delta \bar{x}^\mu}{\delta\varpi _{a}}+\frac{\delta
\bar{x}^\mu}{\delta\varpi _{a}}\star\varpi
_{a})\star\partial_{\mu}\mathcal{Q}_{h}(t,\bar{x})+\textbf{O}(\varpi^{2}).
\end{eqnarray}
The group element $g=e_{\star}^{-\frac{i}{2}\{\varpi _{a},
G_\mu^{a}\}_{\star}}\in{ U_\star(N)}$, where ${ U_\star(N)}$ is
the NC gauge group.
The  noncommutative generators $G_\mu^{a}$ are determined by the
relation:
\begin{eqnarray}
\frac{i}{2}\Big\{\varpi _{a}, G_\mu^{a}\Big\}_{\star}\star
\mathcal{Q}_{h}(t,\bar{x})&=&\frac{1}{2}\Big\{\frac{\delta
\bar{x}^{\mu}}{\delta \varpi _{a}},\varpi
_{a}\Big\}_{\star}\star\partial
_{\mu}\mathcal{Q}_{h}(t,\bar{x})\cr &&-\frac{1}{2}\Big\{\varpi
_{a},\frac{\delta \mathcal{F}(\mathcal{Q}_{h}(t,\bar{x}))}{\delta
\varpi _{a}}\Big\}_{\star}.
\end{eqnarray}
Let us now write  the nonlocal Lagrangians (\ref{11}) in the following form:
\begin{eqnarray}\label{lnc}
\mathcal{L}_{h}(t,\bar{x})&=&\mathcal{L}_{h}^\star(\mathcal{Q}_{h}(t,\bar{x}),\partial_\mu
\mathcal{Q}_{h}(t,\bar{x}),\bar{x})\\
&=&\label{lc}\mathcal{L}_{h}\Big(\mathcal{Q}_{h}(t,\bar{x}),\partial_\mu\mathcal{Q}_{h}(t,\bar{x}),
\partial_\mu\partial_\nu\mathcal{Q}_{h}(t,\bar{x}),\cdots,
\bar{x};\Theta^{\alpha\beta}\Big).
\end{eqnarray}
Remark that in equation (\ref{lnc}), all products are the star
ones and the EL equation of motion can be written in a similar
form as in the usual commutative field theories:
\begin{eqnarray}
\frac{\partial
\mathcal{L}_{h}^\star}{\partial\mathcal{Q}_{h}}-\partial_{\mu}\frac{\partial
\mathcal{L}_{h}^\star}{\partial\partial_{\mu}\mathcal{Q}_{h}}=0.
\end{eqnarray}
Setting  $\zeta(\varpi, f)=\frac{1}{2}\Big(\varpi
_{a}\star\frac{\delta f}{\delta\varpi _{a}} +\frac{\delta
f}{\delta\varpi _{a}}\star\varpi _{a}\Big),$  then
$\partial^t_\mu=\Big(\delta_\mu^\nu-\partial_\mu\zeta(\varpi,
\bar{x})\Big)\partial_\nu$ and we can deduce the  identity
$d^{D}\bar{x}'=[1+\partial_{\mu}\zeta(\varpi,
\bar{x})+\textbf{O}(\varpi^{2})]\mbox{d}^{D}\bar{x}$. Using the
relation (\ref{lnc}), a direct evaluation of $\delta\mathcal{S}$
yields
\begin{eqnarray}
\delta\mathcal{S}&=&\mathcal{S}^t-\mathcal{S}\cr &=&\int
dtd^{D}\bar{x}'\,\mathcal{L}_{h}^\star(\mathcal{Q}^t_{h}(t,\bar{x}'),\partial^t_\mu
\mathcal{Q}^t_{h}(t,\bar{x}'),\bar{x}')-\int
dtd^{D}\bar{x}\,\mathcal{L}_{h}^\star(\mathcal{Q}_{h}(t,\bar{x}),\partial_\mu
\mathcal{Q}_{h}(t,\bar{x}),\bar{x})\cr &=&\int\,dt
d^{D}\bar{x}\,\Big[(1+\partial_\mu\zeta(\varpi,
\bar{x}))\star\mathcal{L}_{h}(t,\bar{x})+\zeta(\varpi,
\mathcal{F})\star\frac{\partial\mathcal{L}_{h}(t,\bar{x})}{\partial
\mathcal{Q}_{h}(t,\bar{x})}\cr &&+\zeta(\varpi,
\bar{x})\star\partial_\mu(\mathcal{L}_{h}(t,\bar{x}))+\partial_\mu(\zeta(\varpi,
\mathcal{F}))\star\frac{\partial\mathcal{L}_{h}(t,\bar{x})}{\partial
\partial_\mu\mathcal{Q}_{h}(t,\bar{x})}\cr
&&-\partial_\mu\zeta(\varpi,
\bar{x})\partial_{\nu}\mathcal{Q}_h(t,\bar{x})\star\frac{\partial\mathcal{L}_{h}(t,\bar{x})}{\partial
\partial_\mu\mathcal{Q}_{h}(t,\bar{x})}+\textbf{O}(\varpi^2)\Big]-\int dtd^{D}\bar{x}\,\mathcal{L}_{h}(t,\bar{x})\cr &=&\int
dtd^{D}\bar{x}\,\Big[-\partial^\mu \mathcal{J}_{\mu}^a+\zeta(\varpi,
\mathcal{F})\star\Big(\frac{\partial
\mathcal{L}_{h}}{\partial\mathcal{Q}_{h}}-\partial_{\mu}\frac{\partial
\mathcal{L}_{h}}{\partial\partial_{\mu}\mathcal{Q}_{h}}\Big)+\mathcal{B}(\varpi)\Big].
\end{eqnarray}
In this expression, the first term is a {\it divergence} term defining the NC tensor $\mathcal{J}_\mu^a$ expressed as follows:  
\begin{eqnarray}\label{cur}
\mathcal{J}_{\mu}^a=\frac{1}{4}\Big\{\Big\{\varpi
_{a},\frac{\delta \bar{x}^{\nu}}{\delta\varpi _{a}}\Big\}_{\star},
\mathcal{T}_{\mu\nu}\Big\}_{\star}-\frac{1}{4}\Big\{\Big\{\varpi
_{a},\frac{\delta \mathcal{F} (\mathcal{Q}_{h}(t,\bar{x}))}{\delta
\varpi _{a}}\Big\}_{\star},\frac{\partial\mathcal{L}_{h}
(t,\bar{x})}{\partial\partial _{\mu}\mathcal{Q}_{h}(t,\bar{x})} \Big\}_{\star}.\nonumber\\
\end{eqnarray}
The second term contains the EL equation of motion while the last term, usually called the breaking term, is given by the relation
\begin{eqnarray}
\mathcal{B}(\varpi)=\frac{1}{4}\Big\{\zeta(\varpi,
\bar{x}),\partial_{\mu}\Big(\{\partial_{\nu}\mathcal{Q}_{h}(t,\bar{x}),\frac{\partial\mathcal{L}_{h}(t,\bar{x})}{\partial
\partial_\mu\mathcal{Q}_{h}(t,\bar{x})}\}_\star\Big)\Big\}_\star.
\end{eqnarray}

$\mathcal{T}_{\mu\nu}$ is the energy-momentum tensor computed in
(\ref{tensor}), defined with non-local variables
$\mathcal{Q}_{h}(t,\bar{x})$:
\begin{eqnarray}\label{tensor1}
\mathcal{T}_{\rho\mu}=\frac{1}{2}\{\partial_{\rho}\mathcal{Q}_{h}(t,\bar{x}),\partial_{\mu}\mathcal{Q}_{h}(t,\bar{x})\}_{\star}
-g_{\rho\mu}\mathcal{L}_h^{\star}.
\end{eqnarray}
The translational invariance violation, engendered by the appearance of
the coordinate $\tilde{\bar{x}}^\mu$, can be avoided by imposing
the constraint
\begin{eqnarray}\label{cc}
\pi(\mathcal{Q}(t,\bar{x}),\tilde{x})\equiv\delta
\mathcal{S}(t,\bar{x})/\delta\tilde{\bar{x}}^\mu \approx 0.
\end{eqnarray}

It is worth noticing that if $\varpi_a\star (\delta
x^\mu/\delta\varpi_a)$ is a constant parameter  and $\mathcal{F}$ is
trivial, then 
the current (\ref{cur}) is reduced to the  NC energy momentum
tensor (\ref{tensor1}).
If $\varpi_a\star (\delta x^\mu/\delta\varpi_a)$ is defined as
$\varpi^{\mu\nu}_a x_\nu,$ where $\varpi^{\mu\nu}_a$ is the Lorentz
tensor and $\varpi_a\star (\delta
\mathcal{F}/\delta\varpi_a)=-\varpi^{\mu\nu}_a
x_\nu\partial_\mu\mathcal{Q}_h(t,\bar{x})$, then the 
current (\ref{cur}) is reduced to the angular momentum tensor.
The current $\mathcal{J}_{\mu}^{a}$ is not symmetric, nonlocally
conserved, and in massless theory, not traceless.

\section{Concluding remarks}
We have provided  a generalization of the  Hamiltonian
formulation developed by Gomis {\it et al} \cite{gommi}, which has
been applied to the renormalizable Grosse-Wulkenhaar $\phi^{4}_{\star D}$ model.
The Euler-Lagrange equation of motion has been derived. The 
 constraints and NC currents have been  investigated and analyzed. The
following statements are worthy of attention:
\begin{enumerate}
\item It is possible to study the original $D$ dimensional non-local Lagrangian system 
(\ref{lgw}) describing the renormalizable GW model as a $D+1$ dimensional local
(in one of the times) Hamiltonian system, governed by the
Hamiltonian  (\ref{tha}) and a set of constraints.
\item Examples of  Hamiltonian symmetry generators of class of  the renormalizable GW model working in a $D+1$ dimensional space
can be given.
\item As expected from previous investigations on NC Noether currents, the tensor $\mathcal{J}_{\mu}^{a}$  (\ref{cur}) is not symmetric,
nonlocally conserved, and, in massless theory, not traceless.
\item A characteristic feature of the Hamiltonian formalism for non-local theories is that it contains the Euler-Lagrange equations as Hamiltonian constraints.
The Euler-Lagrange equation of motion is a constraint in the space
of trajectories.
\end{enumerate}
The EL equation of motion  in $D+1$ dimensions can be also solved
using the  matrix base formalism. In that case, the matrix
elements  can be written as:
\begin{eqnarray}
\mathcal{B}_{h,kl}^{(D+1)}(t,\bar{x})=\int\,
dt'\,\omega_h(t-t')e^{t\frac{d}{d\bar{x}}}\Big(b_{kl}^{(D)}(\bar{x})\Big).
\end{eqnarray}
where $e^{t\frac{d}{d\bar{x}}}$ can be taken as
the evolution operator $T_t$ (translation operator).
The fields $\mathcal{Q}_{h}(t,\bar{x})$ can be reexpressed as
follows:
\begin{eqnarray}
\mathcal{Q}_{h}(t,\bar{x})=\sum_{k,l}C_{kl}\mathcal{B}_{h,kl}^{(D+1)}(t,\bar{x}).
\end{eqnarray}
Then, the formalism developed in  \cite{Gracia-Bondia} can be
applied step by step. Further, the same matrix base method can be adapted to
formulate the NC tensors $\mathcal{J}_\mu^a$. Unfortunately, such
a  computation is too tedious and gives rise to cumbersome
expressions that  are irrelevant for this work. Moreover, their
interpretation needs more investigations whose results will be in
the core of  a forthcoming paper.

\section*{Acknowledgments}
This work is partially supported by the Abdus Salam International
Centre for Theoretical Physics (ICTP, Trieste, Italy) through the
Office of External Activities (OEA) - ICMPA - \mbox{Prj-15}. The
ICMPA is in partnership with the Daniel Iagolnitzer Foundation
(DIF), France. The authors thank the referees for their useful
comments which helped them to improve  the paper. 

\noindent


\begin{thebibliography}{9}
\bibitem{doplicher}
S. Doplicher,  K. Fredenhagen and J. E.Roberts, {\it  The
quantum structure of spacetime at the Planck scale and quantum
fields},  {\it Comm. Math. Phys.} {\bf Vol.~no.~172},
pp.~187-220, (1995).
\bibitem{frank}
P. Aschieri, C. Blohmann, M. Dimitrijevic,   F.  Meyer, P. Schupp,  and J. Wess,
 {\it A gravity theory on noncommutative spaces}, {\it
Class. Quantum Grav.}  {\bf Vol.~no.~22} 3511, (2005).
\bibitem{gommi}
J. Gomis, K. Kammimoura, J. Llosa, {\it Hamiltonian formalism for
space-time non-commutative theories}, Phys. Rev. D {\bf{63}}
(2001), 045003; ArXiv: hep-th/0006235.
\bibitem{llosa}
J. Llosa and J. Vives, {\it Hamiltonian formalism for nonlocal
Lagrangians}, J. Math. Phys. {\bf 35} 2856 (1994).
\bibitem{Douglas}
M. R. Douglas and N. A. Nekrasov, {\it Noncommutative field
theory}, Rev. Mod. Phys. \textbf{73}, 977-1029 (2001); [{\it
e-print} hep-th/0106048]. R. J. Szabo, {\it Quantum field theory
on noncommutative spaces}, Phys. Rept. \textbf{378}, 207-299
(2003); [{\it e-print} hep-th/0109162].
\bibitem{Grosse}
H. Grosse and R. Wulkenhaar, {\it Renormalization of
$\phi^{4}$-theory on non commutative $\mathbb{R}^{4}$ in the
matrix base}, Comm. Math. Phys. \textbf{256}, 305-374 (2005);
[{\it e-print} hep-th/0401128].
\bibitem{llosa2}
J. Llosa, {\it Comment on canonical formalism for Lagrangians
with non-locality of finite extent}, arXiv: hep-th/0201087.
\bibitem{wess} J. Wess,
{\it Deformed coordinate spaces, derivatives}, {\it unpublished},
[e-print {\tt hep-th/0408080}]; {\it ibid}, {\it Deformed gauge
theories}, {\it unpublished}, [e-print {\tt hep-th/0608135}].

\bibitem{chai}
M. Chaichian, P.~P. Kulish, K. Nishijima and A. Tureanu, {\it On a
Lorentz-invariant interpretation of noncommutative space-time and
its implications on noncommutative QFT}, {\it Phys. Lett. B}
\textbf{604}, 98--102 (2004); [e-print {\tt hep-th/0408069}]. F.
Koch and E. Tsouchnika, {\it Construction of $\theta$-Poincar\'e
algebras and their invariants on $M_\theta$}, {\it Nucl. Phys. B}
\textbf{717}, 387--403 (2005); [e-print {\tt hep-th/0409012}]. J.
Lukierski, A. Nowicki and H. Ruegg, {\it Phys. Lett. B}
\textbf{293}, 344--352 (1992).
\bibitem{LS}
E. Langmann and R.J. Szabo, {\it Duality in scalar field theory
on noncommutative phase spaces}, {\it Phys. Lett. B} \textbf{533},
168--177 (2002); [e-print {\tt hep-th/0202039}].
\bibitem{Gerh}
A. Gerhold, J. Grimstrup, H. Grosse, L. Popp, M. Schweda and R.
Wulkenhaar, {\it The energy momentum tensor on noncommutative
spaces - some pedagogical comments\/}, {\it unpublished}, [e-print
{\tt hep-th/0012112}].

\bibitem{Abu}
M. Abou-Zeid and H. Dorn, {\it Comments on the energy momentum
tensor in noncommutative field theories\/}, {\it Phys. Lett. B}
\textbf{514}, 193-188 (2001); [e-print {\tt{hep-th/0104244}}].

\bibitem{Grims}
J.M. Grimstrup, B. Kloib\"ock, L. Popp, V. Putz, M. Schweda and
M. Wickenhauser, {\it The energy momentum tensor in noncommutative
gauge field models\/}, {\it  Int. J. Mod. Phys. A} \textbf{19},
5615--5624 (2004); [e-print {\tt hep-th/0210288}].
\bibitem{Ben-Houk1}
J. Ben Geloun and  M. N. Hounkonnou, {\it  Energy momentum tensors
in  renormalizable noncommutative scalar field theory}, Physics.
Letter B \textbf {653}, 343-345 (2007).
\bibitem{woodard}
R. P. Woodard, {\it A canonical formalism for lagrangians with
nonlocality of finite extent}, hep-th/0006207.
\bibitem{Ben-Houk2}
J. Ben Geloun and  M. N. Hounkonnou,{\it Noncommutative Noether
theorem}, AIP Proc \textbf{956} 55-60 (2007).
\bibitem{Dmitri}
Dmitri V. Vassilevich, {\it Constraints, gauge symmetries, and
noncommutative gravity in two dimensions}, hep-th/0502120.
\bibitem{Toni}
J. Gomis, K. Kamimura, T. Mateos, {\it Gauge and BRST generators
for space-time non-commutative $U(1)$ theory}, hep-th/0009158.
\bibitem{Gracia-Bondia}
J.M. Gracia-Bond\'ia,  and J.C. V\'arilly,   {\it Algebras of
distributions suitable for phase-space quantum mechanics I} {\it
J. Math. Phys.} {\bf Vol.~no.~29}, pp.~869-879, (1988).
\bibitem{Varilly}
J.M. Gracia-Bond\'ia,  and  J.C. V\'arilly,  {\it Algebras of
distributions suitable for phase-space quantum m\'ecanics II.
Topologies on the Moyal algebra}, {\it J. Math. Phys} {\bf
Vol.~no.~29}, (1988).
\bibitem{Dirac}
P. A. M. Dirac, {\it The Fundamental equations of quantum
mechanics}, Proc.
    Roy. Soc. Lond. A 109, 642 (1925);
{\it On quantum algebra}, Proc. Cambridge
    Phil. Soc. 23, 412 (1926).
\bibitem{Rivasseau}
V. Rivasseau, {\it Noncommutative renormalization}, S\'eminaire
Poincar\'e X Espace Quantique, Inst. Henri Poincar\'e, Paris,
2007,  15-95.
\bibitem{Goursac}
A. de Goursac,   A. Tanasa, and J.C. Wallet,   {\it Vacuum
configurations for renormalizable noncommutative scalar models},
{\it Eur. Phys. J.C\/} {\bf Vol.~no.~53,459}, (2008).
\end{thebibliography}
\end{document}